\documentclass[structabstract]{aa}  

\usepackage{graphicx}
\usepackage{txfonts}
\begin{document}
   \title{Typical duration of good seeing sequences at Concordia}


   \author{E. Fossat
          \inst{1}
          \and
          E. Aristidi\inst{1}
	  \and A. Agabi\inst{1}
	  \and E. Bondoux\inst{1}
 	  \and Z. Challita\inst{1}
  	  \and F. Jeanneaux\inst{1}
   	  \and D. M\'ekarnia\inst{1}
}

   \institute{Fizeau Laboratory, University of Nice Sophia Antipolis, Parc Valrose, F - 06108 Nice cedex\\
              \email{eric.fossat@unice.fr}}

   \date{Received September 15, 1996; accepted March 16, 1997}

 
  \abstract
   {The winter seeing at Concordia is essentially bimodal, excellent or quite poor, with relative proportions that depend on altitude above the snow surface. This paper studies the temporal behavior of the good seeing sequences.}
   {An efficient exploitation of extremely good seeing with an adaptive optics system needs long integrations. It is then important to explore the temporal distribution of the fraction of time providing excellent seeing.}
   {Temporal windows of good seeing are created by a simple binary process. Good or bad. Their autocorrelations are corrected for those of the existing data sets, since these are not continuous, being often interrupted by technical problems in addition to the adverse weather gaps. At the end these 	corrected autocorrelations provide the typical duration of good seeing sequences. This study has to be a little detailed as its results depend on the season, summer or winter.}
   {Using a threshold of 0.5 arcsec to define the ``good seeing'', three characteristic numbers are found to describe the temporal evolution of the good seeing windows. The first number is the mean duration of an uninterrupted good seeing sequence: it is $\tau_0=7.5$~hours at 8~m above the ground (15 hours at 20~m). These sequences are randomly distributed in time, with a negative exponential law of damping time $\tau_1=29$~hours (at elevation 8~m and 20~m). The third number is the mean time between two 29~hours episodes. It is $T=10$~days at 8~m high (5 days at 20~m).
}
   {There is certainly no other site on Earth, except for the few other high altitude Domes on the Antarctic plateau, that can get close to these really peculiar seeing conditions.}

   \keywords{Antarctica --
                Site testing --
     }

   \maketitle
%

\section{Introduction}

Regarded as astronomical sites,  the highest points of the Antarctica plateau present many obvious advantages due to the local climate and the remoteness from any polluting civilization. They also benefit from an interestingly unique distribution of turbulence. This has been extensively measured at Dome C since the first winter-over permitted in 2005 by the French - Italian Concordia station operation (Aristidi et al, 2009). Winter and Summer display very different but both unusual vertical distributions of the turbulent energy. Generally speaking, the situation is dominated by the presence of a surface inversion layer that becomes very turbulent when the temperature gradient is strong in winter, and can completely vanish in summer when this gradient becomes flat. In summer it depends on the Sun's elevation, and is then strongly time dependent, with an optimum period of a few hours of excellent seeing every day in the middle of local afternoon (Aristidi et al, 2005). In the other 3 seasons, the mean seeing is essentially altitude dependent above the snow surface. Above this surface turbulent layer that contains, statistically, 95 percent of the total C$n^2$ along the line of sight, it has proved to be statistically independent of the season, within the measurements accuracy. Its mean value is between 0.3 and 0.4 arcsec as soon as the telescope is located above a sharply defined altitude threshold, that fluctuates around a mean value of the order of 25 m. The consequence is that the non summer seeing displays a nearly bimodal statistical distribution. It is indeed as good as 0.3 to 0.4 arcsec 50 percent of the time at 25m above the surface, this fraction of time decreasing to about 40  percent at 20m and slightly less than 20 percent at 8m. But it is obviously not equivalent to have 40 percent of good seeing spread in many short sequences of seconds to minutes rather than distributed in extended long sequences measured in hours or days. This paper addresses the temporal distribution of this good seeing percentage. It goes beyond the first analysis made in Aristidi et al. (2009), using a method to compensate for the gaps in the data.


\section{The autocorrelation method}

This paper exploits the Concordia DIMM data. A DIMM located at about 8m above the snow surface is permanently operated since the end of 2004, after having already provided a summer season one year before. A second DIMM was set at 20m high on the roof of the Concordia station during 3 months in 2005. There is also a GSM, made of two DIMMs on the snow surface. They are not exploited in the present paper.

The DIMM seeing data sets have a sampling rate of 2 minutes. Their data time series are not continuous, being relatively often interrupted either by adverse weather, or by various technical reasons such as frost on the optics, electronics or computer shut downs, loss of star tracking, .... It is then difficult to directly track the continuity of good seeing sequences in the original data files as a gap in such a sequence of good seeing data can be due either to a bad seeing moment or to a simple lack of data. This study must then be made on a statistical basis. 

The autocorrelation method applied to data windows proved to be a very efficient tool for that purpose. Two different temporal windows can be defined, by means of a function equal to 1 or 0.

The first one is called $w_e$(t) and defines the existence of data at a time $t$. $w_e$(t)=1 if data exist, 0 otherwise.

The second one $w_g$(t) defines the seeing quality. It uses a threshold $\sigma_g$ so that $w_g$(t) =1 if the seeing $\sigma(t)<\sigma_g$ and $w_g$(t)=0 if $\sigma(t)>\sigma_g$. In all the rest of this paper, we will use $\sigma_g$ = 0.5~arcsec, a value that is near the minimum of the gap between good and bad seeing in the winter seeing histograms (Aristidi et al, 2009).

This second window function, however, is not directly accessible since a significant fraction of measurements is missing. The only accessible one is the product of these two windows, that will be 1 when the measurement does exist and provides a good seeing value, and 0 when either the seeing is not good or the measurement does not exist. This accessible window function can be called                  $w_{ge}$(t)=$w_e$(t) $w_g$(t)

It is unfortunately not possible to recover the interesting but unknown window $w_g$(t) as a division by $w_e$(t) that contains many zeroes is not feasible. Such a division, however, is possible in the domain of the autocorrelations, as long as the data sets are long enough regarding their characteristic time. Assuming that $w_g$ and $w_e$ are statistically independent, the autocorrelation of $w_{ge}(t)$ can be written as
\begin{eqnarray}
\Gamma_{ge}(\tau)&=\frac{1}{m_e\,m_g} E[w_e(t)\,w_g(t)\,w_e(t+\tau)\, w_g(t+\tau)]\\
                 &= \frac{1}{m_e\,m_g} E[w_e(t)\,w_e(t+\tau)]\: E[w_g(t)\,w_g(t+\tau)] \nonumber\\
                 &= \Gamma_e(\tau)\: \Gamma_g(\tau) \nonumber
\end{eqnarray}
where $\Gamma_e(\tau)$, $\Gamma_g(\tau)$ and $\Gamma_{ge}(\tau)$ are the autocorrelation functions of $w_e$(t), $w_g$(t) and $w_{ge}$(t) respectively, $m_e$ and $m_g$ are the mean values of $w_e$ and $w_g$ (the mean value is here equal to the variance since the functions $w_e$ and $w_g$ equal 0 or 1) and $E[]$ denotes the expected value. This technique has been extensively used in the past to analyse heliosismic data (Lazrek, 1993) and is in fact a method of deconvolution in the Fourier plane. As we benefit of very long time series, spread over several years of data that can be cumulated, and as  there is very little suspicion of dependence between the observation window $w_e$(t) and the good seeing occurrences $w_g$(t), we can use this division and thus study the statistical properties of the good seeing sequences through the autocorrelation $\Gamma_g(\tau)$ of the corresponding window $w_g$(t).

It is now well established that the summer and winter conditions at Concordia are drastically different. We are mostly interested in this paper by the winter conditions. The "winter" in this study will be defined as the 6 months during which no significant temperature variation is visible, which is from early April to the end of September. We will also qualify and illustrate the method by a summer analysis,  the summer data being taken during the permanent sunlight season, from early November to February 10th.

Figures 1 and 2 illustrate the data window function autocorrelation $\Gamma_e(\tau)$. During the first two winters, in 2005 and 2006, only one astronomer was present on the site. It is clear on Fig. 1 that the automation of the instrument was not fully successful yet, as demonstrated by the one-day periodicity of the window function. In 2007 for the first time, two astronomers wintered over and could then take care of the instrument more permanently. The one-day periodicity of the window function is not totally cancelled, but almost. 

The overall filling factors of the data sets are provided by the asymptotic values of the autocorrelations functions. It is indeed well known (Papoulis, 1984) that $\Gamma_e(\tau)\rightarrow E[w_e(t)]^2/m_e=m_e$ where $\tau\rightarrow\infty$, as the quantities $w_e(t)$ and $w_e(t+\tau)$ become statistically independent. The autocorrelation tends towards the mean value of $w_e$, i.e. the time percentage where $w_e=1$. This filling factor was found to be about 40 percent during the first two years and slightly more than 30 percent in 2007, due to a relatively long technical interruption during the first half of that winter season.

\begin{figure}
\centering
\includegraphics[bb=60 270 580 570,width=\columnwidth]{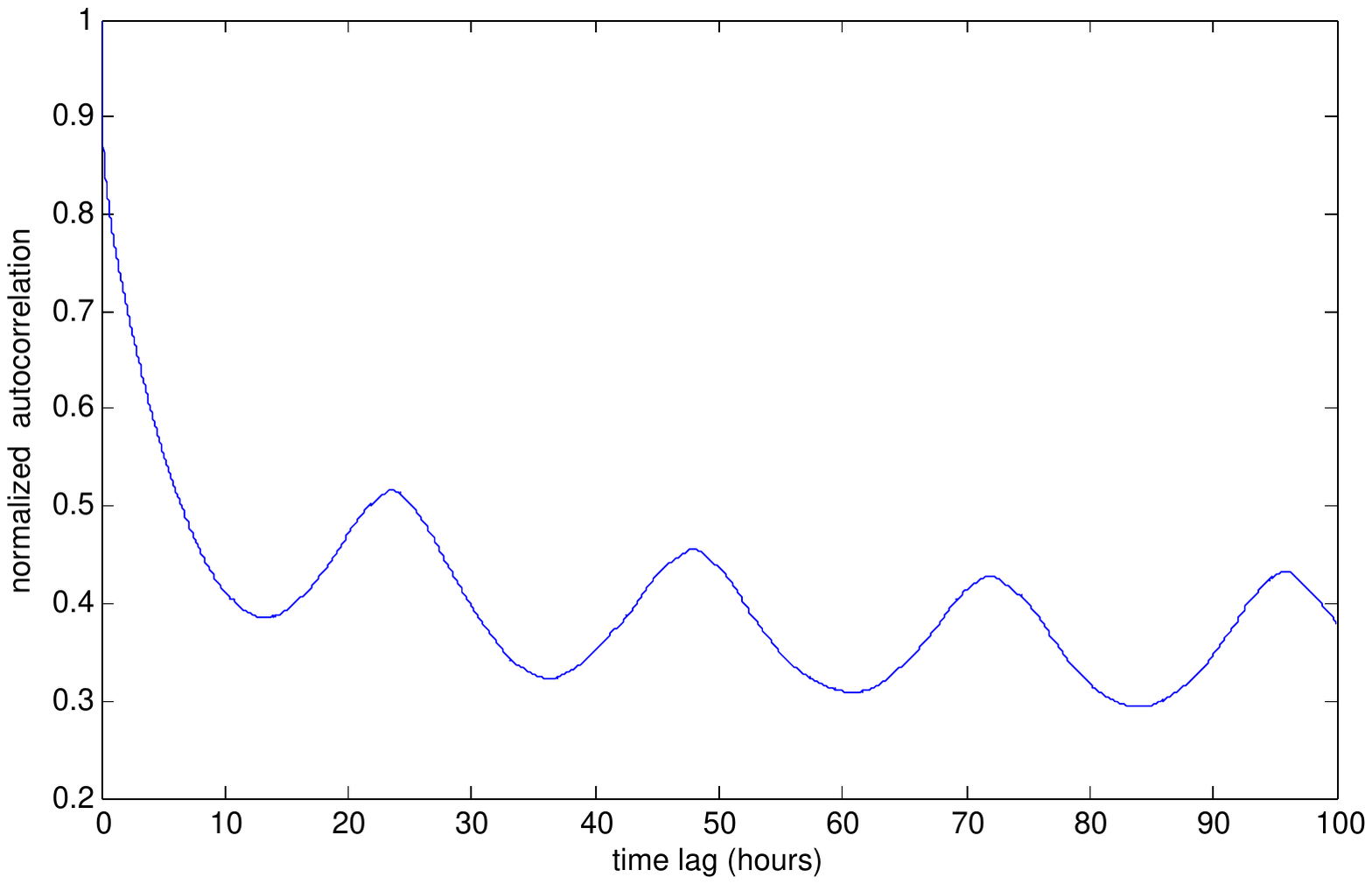}
\caption{Autocorrelation $\Gamma_e(\tau)$ of the existing data window, averaged on the 2005 and 2006 winter seasons. The one-day periodicity is clearly visible.}
\label{default}
\end{figure} 

\begin{figure}
\centering
\includegraphics[bb=60 270 580 570,width=\columnwidth]{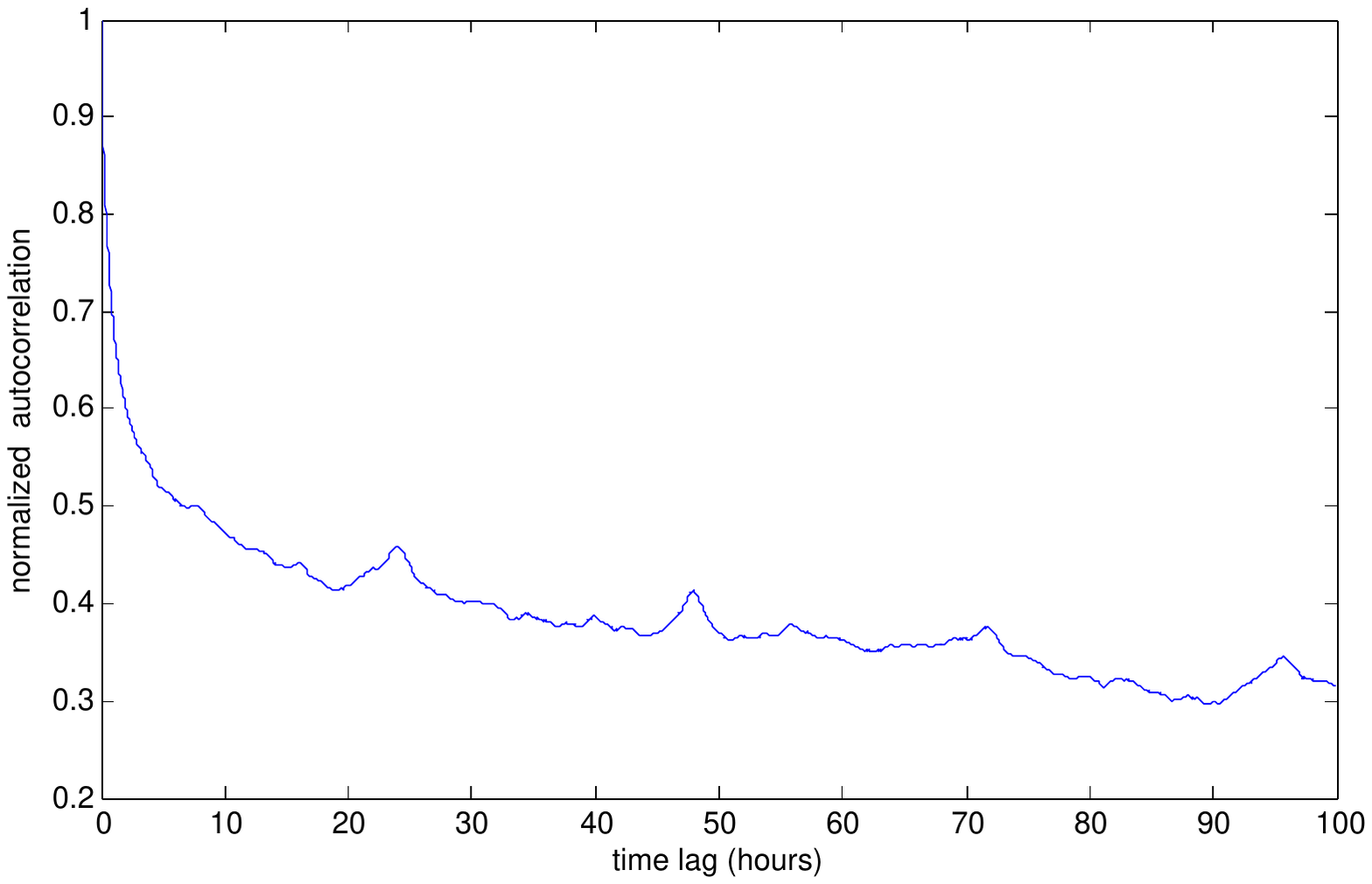}

\caption{Autocorrelation $\Gamma_e(\tau)$ of the winter 2007 data window function. A long gap due to an important technical problem occurred during that winter, so that the overall filling factor is a little less, but with two astronomers wintering over instead of only one, the amplitude of the 
24-hour periodicity is drastically reduced.}\label{default}
\end{figure} 

\section{The summer situation}

In summer, there are generally a few more observers, but there is also more technical maintenance, so that the gaps in the data are not less than in winter. The overall filling factors are still in the range 30 to 40 percent but with several people taking care, the one-day periodicity is almost absent in the window $w_e$(t) (not shown here). However, the seeing itself has a 24-hour periodicity as it displays a minimum every day in the afternoon (Aristidi et al, 2005), so that the autocorrelation of the good seeing window function  $\Gamma_g(\tau)$, obtained by the division of  $\Gamma_{ge}(\tau)$ by  $\Gamma_e(\tau)$, is then expected to display a large amplitude one-day periodicity. That is confirmed by Fig.3, averaged on all available summer seasons from 2003/2004 up to 2007/2008 and then statistically very robust.

\begin{figure}
\centering
\includegraphics[bb=30 260 590 570,width=\columnwidth]{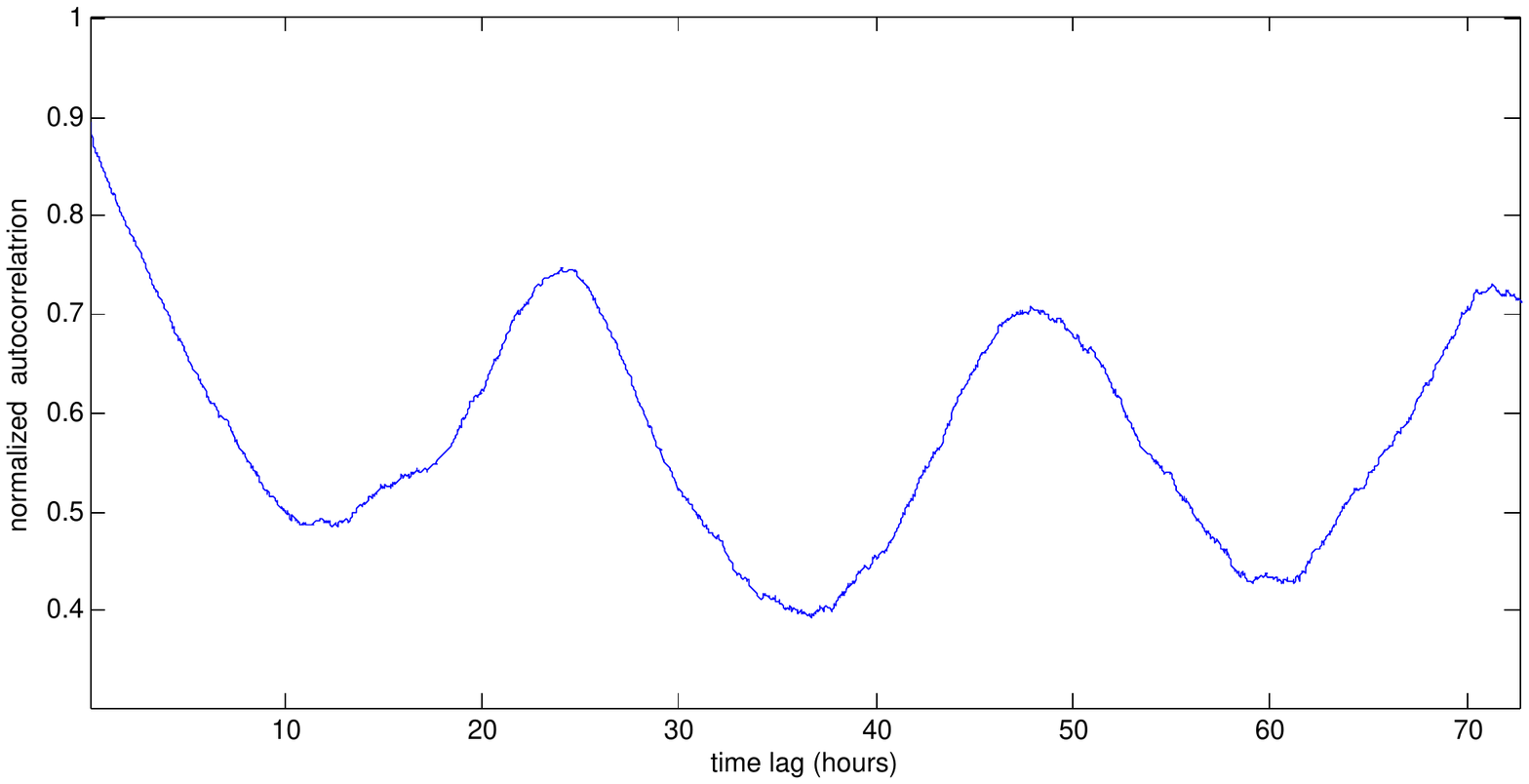}
\caption{Autocorrelation $\Gamma_g(\tau)$ of the good seeing data window, averaged on all the summer seasons from 2003/2004 up to 2007/2008. This time, the one-day periodicity is not due data gaps but to a real daily variability of the seeing.}
\label{default}
\end{figure} 

This figure deserves some comments. The first comment is about the asymptotically behaviour of the autocorrelation, which oscillates between 0.42 and 0.72, around a mean value of 0.57. It means that 57 percent of the time during the 3-month summer season the seeing is better than our threshold $\sigma_g$ =  0.5 arcsec. Another comment concerns the first 6 hours, where the autocorrelation shows an almost exactly linear decrease. Such a linear decrease corresponds to a rectangular window function. It thus means that a significant part of the summer good seeing sequences is indeed made of continuous sequences of 6 hours long on average, every day at the same time. This is exactly what we already know of the summer seeing (see Fig. 3 in Aristidi et al, 2005), which has been measured to be better than 0.5 arcsec between 2 and 8 p.m. every day in summer. 

It can be noted that 6 hours per day represents 25 percent, while the total fraction of good seeing data is statistically more than twice this percentage. There are then many other episodes of good seeing in summer. They are shorter and more randomly distributed. For instance, the quick decrease of the autocorrelation from 1 to $\sim 0.9$ in the first few minutes indicates that an order of 10 percent of good seeing or bad seeing events occur as isolated events of one or a very few consecutive 2-minute measurements.

Besides that last remark, this autocorrelation analysis of the summer data does not provide many new informations indeed. However it does qualify the method and will be used as a reference for the winter data understanding.

\section{The winter situation}

Let's start this section by looking at Fig. 4, the mean autocorrelation function $\Gamma_g(\tau)$ of the good seeing windows, averaged on 4 winter seasons from 2005 to 2008. This is the seeing measured on the Concordiastro platform at 8 m. Thanks to the 24 months of data exploited here, this autocorrelation is statistically very robust again.

\begin{figure}
\centering
\includegraphics[bb=80 270 550 570,width=\columnwidth]{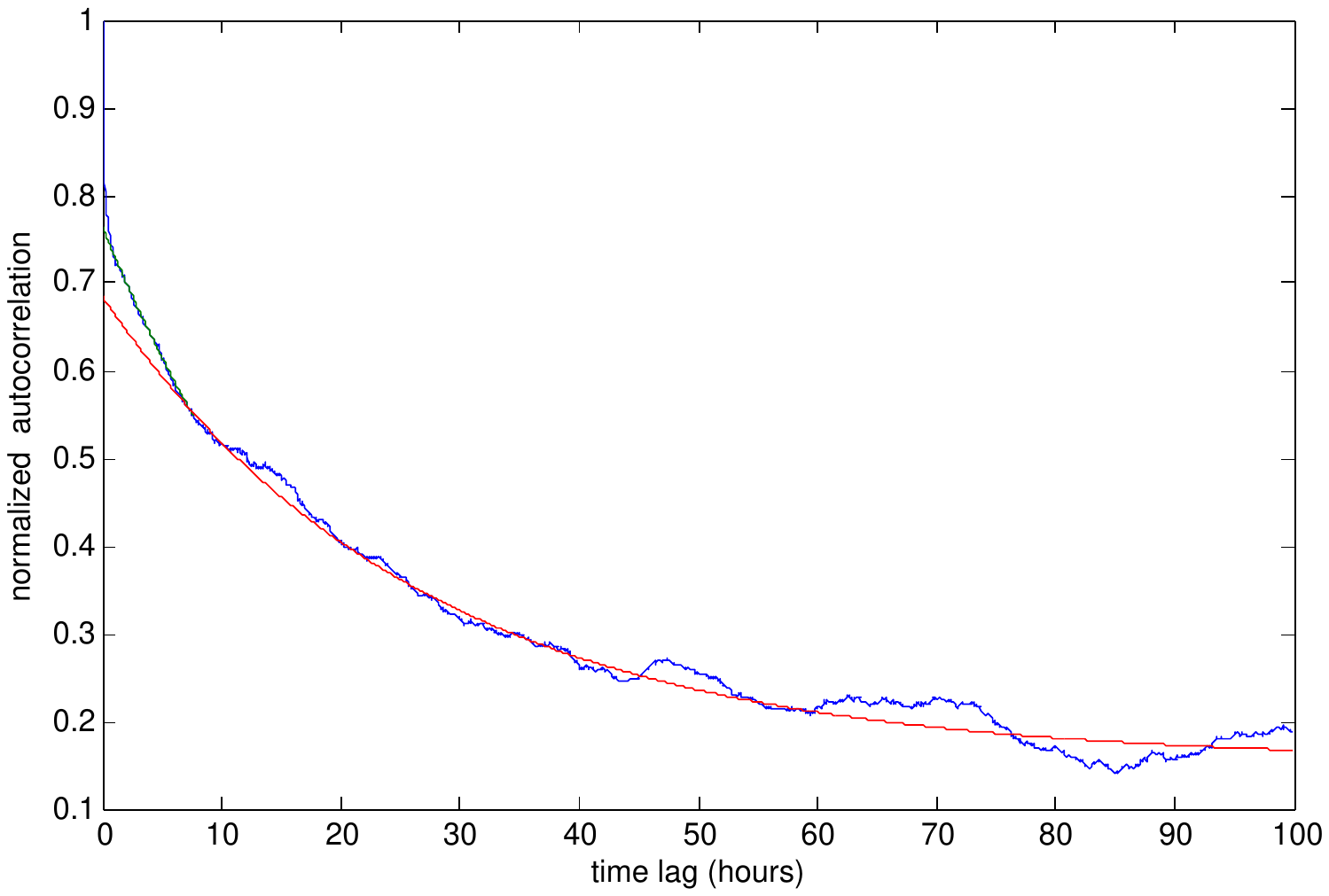}
\caption{Autocorrelation $\Gamma_g(\tau)$ of the good seeing windows, averaged on 4 winter seasons seasons from 2005 to 2008.
The 5 parameters fit described by Eq.~2 is superimposed to the curve.}
\label{default}
\end{figure} 

The asymptotic value, slightly less than 0.2, is a good approximation of the probability of obtaining good seeing values, i.e. being above the turbulent boundary layer at 8 m high. It confirms the estimation of 18\% made by means of the histogram integrals (Aristidi et al, 2009).

The behavior at the origin shows a quick drop from 1 to about 0.8. It confirms the existence of very short sequences (lasting typically less than 10~mn) where the seeing is either continuously good or bad. These individual events were noticed by the observers, and represent 20\% of the data. Apart these first points, the graph of $\Gamma_g(\tau)$ shows three main features. A linear decrease in the first 7~hours, a negative exponential distribution of characteristic time of a few tens of hours, and an horizontal asymptot for large values of $\tau$. We adjusted to $\Gamma_g(\tau)$ the following 5 parameters fit :
\begin{equation} 
C_g(\tau) = C \left(1 - \frac{\tau}{\tau_0}\right)\prod\left( \frac{\tau}{2\tau_0}\right)+B e^{- \frac{\tau}{\tau_1}}+A
\end{equation}
The first term accounts for the linear part and is truncated by the rectangle function $\prod\left( \frac{\tau}{2\tau_0}\right)$ equalling to 0 when $\tau>\tau_0$. We found:
\begin{itemize}
\item The asympotic value $A=0.16$. It represents the overall probability to observe a good seeing
\item $B=0.52$ and $\tau_1=29$ hours, the parameters of the exponential decrease
\item $C=0.08$, $\tau_0=7.5$ hours, the parameters of the linear decrease
\end{itemize}
The linear part is similar to what is obtained in summer. It is indeed the autocorrelation of a rectangular window function. It shows that many good seeing sequences occur in continuous runs lasting $\tau_0$ = 7.5 hours on average. An important difference, though, is that in summer, such sequences of 6 hours reproduce every day at the same time, while in winter, there no significant daily periodicity. The exponential decrease that follows this initial linear behavior is the typical winter feature and must be interpreted.

We propose the following model for the temporal window function $w_g(t)$. This model described hereafter depends on several parameters which were adujsted so that its autocorrelation reproduces correctly the autocorrelation $C_g(\tau)$ of the data.  $w_g(\tau)$ is modeled by well separated episodes including several rectangular functions of width 7.5 hours. The mean delay between two consecutive episodes is noted $T$, with $T=10$~days.

Inside an episode the rectangular functions start at times $t_i$, where $t_i$ is a random variable with a negative exponential distribution of damping time $\tau_1=29$~hours. That means that most of the windows appear at the beginning of the episode and tend to disappear after a delay of 29~hours from the beginning of the episode. The mean number $N_r$ of rectangular windows inside a 10~days episode is adjusted to account for the value of $B$ in the function $C_g(\tau)$ (eq.~2). We find $N_r=11$. These 11 windows are mostly concentrated in the first 29~hours of the episode and there can be large overlaps. In our simulation, more than 75~\% of the good seeing values are concentrated in the first 48~hours of a 10-day episode. Fig. 5 shows an example of such a 2-day episode. 

In addition to these episodes, and to account for the initial quick drop of $\Gamma_g(\tau)$, we simply simulated the initial drop by adding a random number $N_i$ of isolated either good or bad seeing (lasting two minutes).  $N_i$ is adjusted to 0.22.

The general figure at 8 m is then that every 10 days on average, there is a two to three day episode during which many nearly uninterrupted runs of good seeing occur with individual durations of 7.5 hours.

\begin{figure}
\centering
\includegraphics[bb=50 260 580 570,width=\columnwidth]{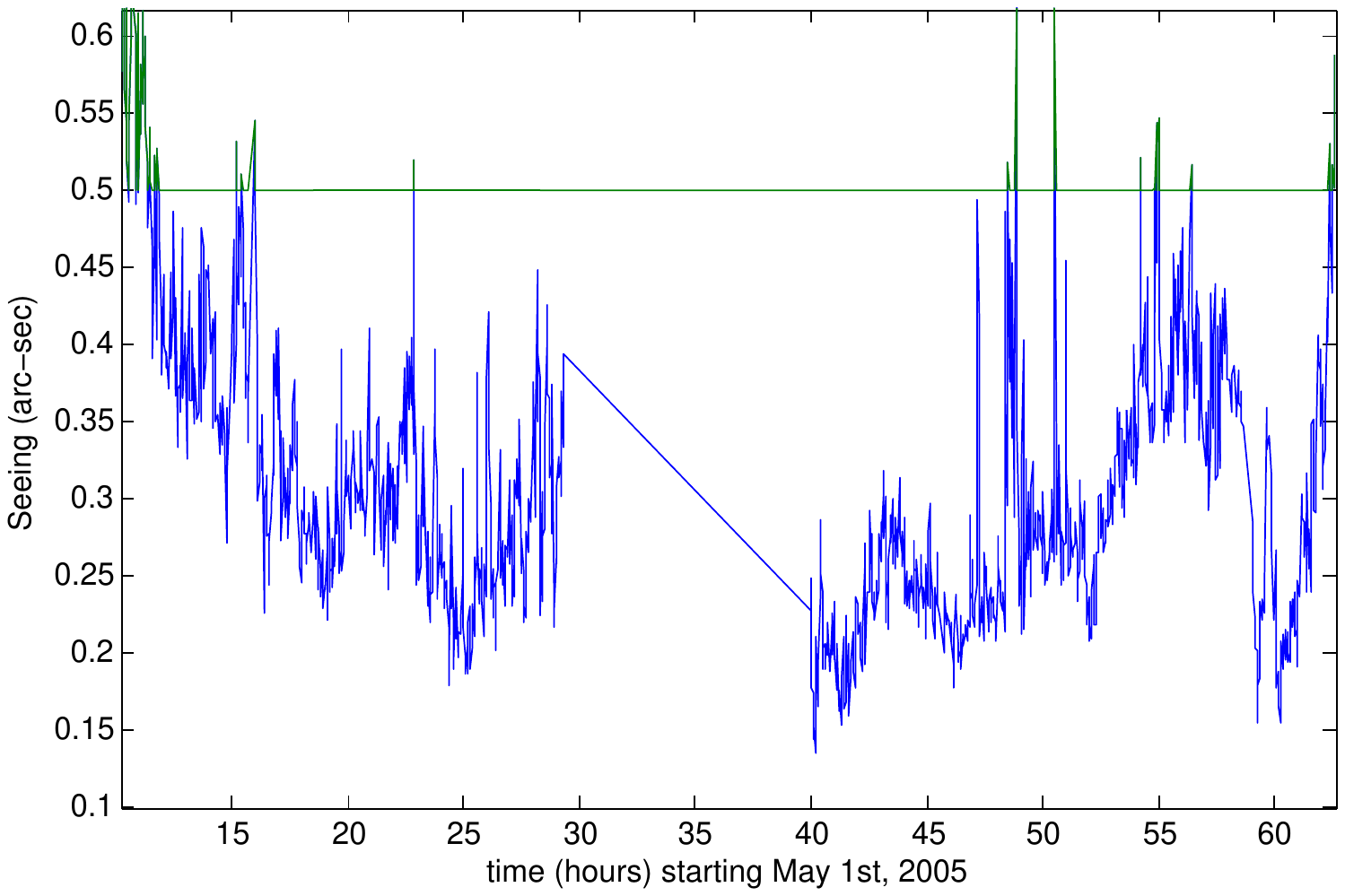}
\caption{A nice sequence of good seeing data obtained in winter 2005 at 8~m high. During 50 hours, there are only a few nearly individual values above the 0.5 arc-sec threshold, but also an 11-hour gap.}
\label{default}
\end{figure} 

\paragraph{The seeing a 20~m high.}
Only a little more than 3 months of data have been collected at this height so that the statistical robustness is not as good as it is at 8~m. Fig. 6 shows the autocorrelation function $\Gamma_g(\tau)$ of the good seeing windows at 20 m.  This curve is clearly not as smooth as the other one, due to its weaker statistical robustness. The one-day periodicity is visible because a large part of this data set comes from Spring (from July, 23rd to October, 31st) when the Sun is present, only part time but a longer and longer part till the end of the run. However,  the same kind of fit can be attempted and works well. Interestingly enough, the 29-hour damping time of the exponential decrease is not modified by this higher altitude. That may presumably be explained by the fact that it depends mostly on the meteorological general situation, that should not be affected by a few meters difference of altitude. On the other hand, the nearly linear decrease that was explained by uninterrupted runs of 7.5 hours is still present, with now a twice longer duration of 15 hours. The numbers $A$ and $B$ are fitted to 0.36 and 0.27 respectively. The value $A=0.36$ is coherent with the good seeing fraction of 40~\% derived from the histograms (Aristidi et al., 2009). In our simulations, $N_r$ must be adjusted between 6 and 7. This again implies lots of overlaps and a mean fraction of good seeing of about 90 percent during the first 48~hours of a simulated episode. The delay $T$ between consecutive episodes is found twice shorter, around 5 days. 

The two delays between consecutive episodes, 10 and 5 days, imply that during the 100-day winter time (without any sunrise), the mean numbers of such good seeing episodes must be about 10 at 8 m and 20 at 20 m.

Statistically, This difference can be explained by assuming that half the good seeing episodes at 20 m correspond to a boundary layer upper limit that goes down below 20 m but not as low as 8 m,  the other half including situations when it will also spend some time below 8 m. The same 29-hour exponential decrease at both altitudes gives credit to our model and its meteorological interpretation.

 It would not be realistic to attempt more interpretation of that curve, that has to be taken as providing an estimate of what may happen at 20 m high. However, that estimate is important as one may consider that a possible optimal choice for a future high angular resolution instrument at Dome C could be to set it at $\sim 20$~m, together with a well qualified GLAO system.  It would then benefit of a statistics of free atmosphere seeing of about 40 percent of the time during all the dark and cold period, concentrated inside episodes of at least two days with several uninterrupted runs of individual durations being at least 15 hours. A number of the order of 20 such episodes can be estimated to occur during all the dark period. That will definitely permit long integrations in very good seeing conditions.

\begin{figure}
\centering
\includegraphics[bb=50 260 550 570,width=\columnwidth]{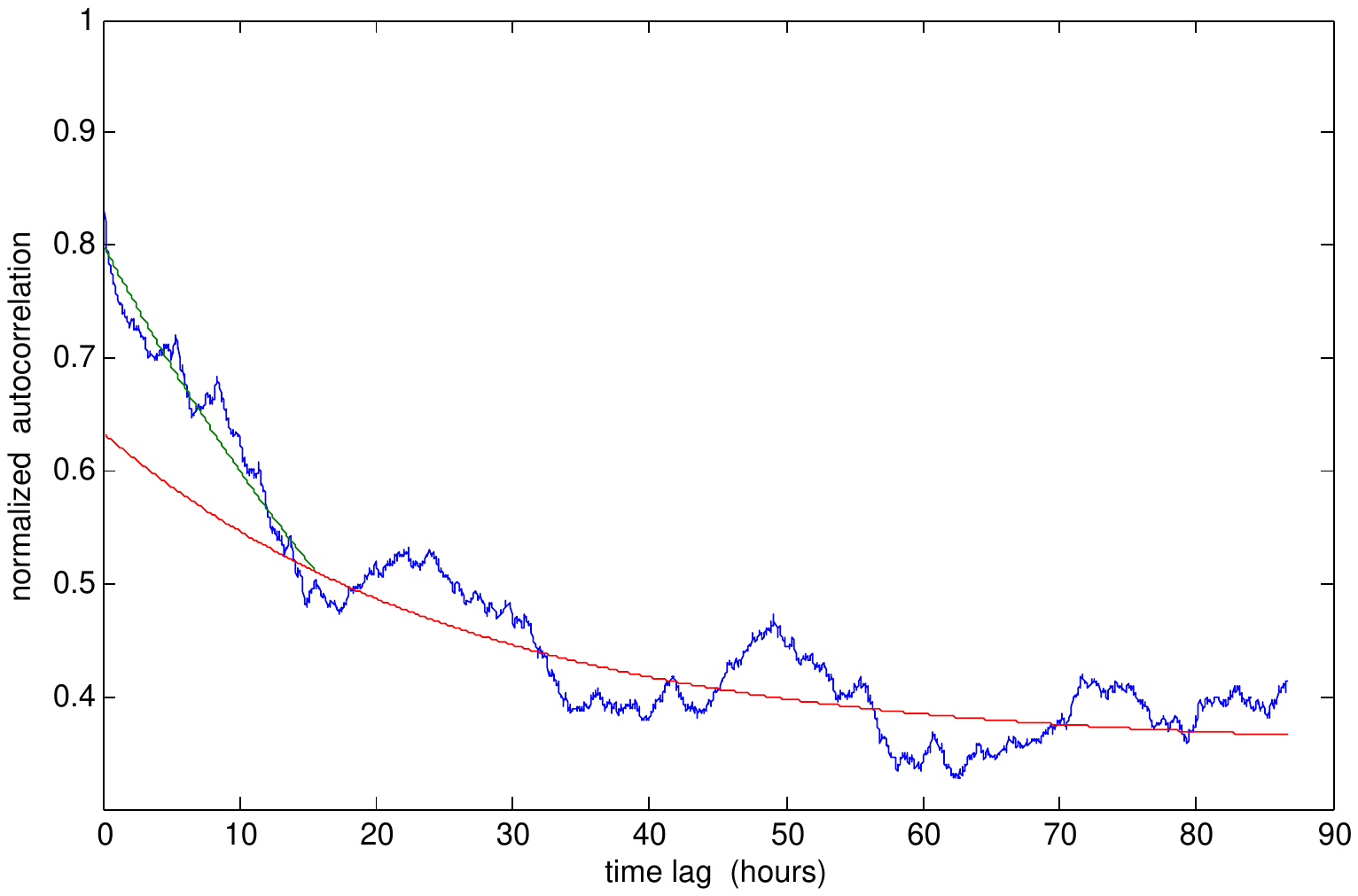}
\caption{Autocorrelation $\Gamma_g(\tau)$ of the good seeing data window, at 20-m high during 3 months in Winter/Spring 2005, with the same kind of fit.}
\label{default}
\end{figure}

\section{Discussion and conclusions}

Thanks to the very long data sets exploited here, the validity of the autocorrelation division for correcting  the data interruptions and the final statistical robustness are well validated by the summer season analysis and results. At least at 8-m high. The reduced amount of data available at 20-m high makes the statistical robustness of the numbers visibly weaker,  but the general tendency tends to reinforce the interpretation of the 8-m data. The 29-hour exponential decay of the distribution of the good runs starting times, assumed to depend on the meteorological situation, is found to be the same at both altitudes. The second characteristic time $\tau_0$, that can be regarded as the minimum duration of nearly uninterrupted good runs, is found twice longer at 20 m, i.e. 15 hours versus 7.5 hours at 8 m. Finally, the number of episodes of excellent seeing is estimated to be twice more frequent at 20m, 20 times versus 10 times per winter, and that is also easy to understand with a reasonable assumption on the vertical motions of the boundary layer upper limit.
    
\begin{acknowledgements}
We wish to thank the French and Italian Polar Institutes, IPEV and PNRA and the CNRS for the financial and logistical support of this programme. Thanks also to the different logistics teams on the site who helped us to set up the experiments. We are also grateful to our industrial partners ``Optique et Vision'' and ``Astro-Physics'' for technical improvements of the telescopes and their help in finding solutions to critical technical failures on the site. Thanks to our colleagues Max Azouit and Fran\c{c}ois Martin for constant help and ideas since the beginning of the project. Thanks to F. Vakili, the director of Fizeau laboratory, for his support. We thank all the people who gave a hand in setting up the instruments during the summer campaigns since the end of 2003. Finally, thanks to the referee who helped improving the presentation.
\end{acknowledgements}

\end{document}